# On the 'Scattered' Inclinations in the Kuiper Belt

Fred A. Franklin and Paul R. Soper
Harvard Smithsonian Center for Astrophysics

Our plan for this paper centers on clarifying what has come to be known as the 'scattered disk' component of the Kuiper belt. This label refers especially to the particular feature that the inclination, i, of its bodies, plotted here in Fig. 1 for the limited semimajor axis range between 38 and 57 AU shows many individual values extending upward to 30, even 35 deg. Entries are the current i's of 712 bodies observed for 3 or more oppositions while Fig. 2 is its counterpart for the eccentricities, e. Figure 3 returns to the i's, concentrating on just those lying within +/- 0.50 AU of the 2/3 mean motion resonance [mmr] at 39.43 AU. It provides a selective and more uniform set of 224 objects, again observed for 3 or more oppositions and one for which the vast majority of those lying there move in very-to-fairly stable orbits, hence avoiding ones soon likely to escape. [A paper (Franklin and Soper, 2014), indicating that most orbits with $0.1 < e < 0.34$ at 2/3 are probably stable over long times, forms the basis for this claim.] A detail worth remembering: current tabulations yield only 'snapshots' of all elements as observations still cover but a fraction of one orbital period.

How to interpret Fig.3 is our chief concern as Fig. 2 proves to be no problem. One might at first suppose that the scattered disk concept would apply to the e's too save for the clear fact that the development of the values shown is not a mystery because much and probably all of the size and range of the e's arises from their substantial growth, after the initial capture over a range of a(o)'s, that is induced by Neptune's migration. This is now a well-investigated process, but the key point is that, though migration will pump the initial e's sometimes even to unstable heights, whether it can also markedly increase the i(o)'s remains decidedly unlikely. We will examine this question using numerical studies in later paragraphs.

Another valuable key comes from the abundance of captured bodies at the 2/3 mmr as it provides a fine selection for sampling the inclination range that arguably might date back to the original i(o)s. Only the mmrs may have the ability, which we shall check, of being able to preserve the distribution in i over times of billions of years. Thus we can squarely face the choice whether the i range of stable objects lying in mmrs of the Kuiper belt reflects either 1) a broad initial distribution of i(o)'s or 2) is the result of a random or evolutionary event that is not directly tied to the migration that raised the e's.

We begin by asking whether a seemingly likely i(o) distribution might fit the observed values. Table I provides this comparison of the observed i's at 2/3 with ones calculated from an integration of 300 bodies, introduced with random i(o) < 8 deg and e(o) < 0.1 of which 53 were captured and retained at 2/3 for at least 1 byr.

**Table I. Observed and calculated populations at the 2/3 mmr. Note: Initially all bodies in the calculated case were randomly chosen with e(o)<0.1, i(o)<8 deg. The final two columns apply after an integration of 1 byr.**

| i range (deg) | Observed No. >3 oppositions | % | Observed No. >4 oppositions | % | Calculated No. | % |
|---|---|---|---|---|---|---|
| 0 - 10 | 115 | 51 | 64 | 49 | 44 | 83 |
| 10 - 20 | 78 | 35 | 50 | 38 | 9 | 17 |
| 20 - 30 | 24 | 11 | 12 | 9 | 0 | 0 |
| > 30 | 7 | 3 | 4 | 3 | 0 | 0 |
| Totals | 224 | | 130 | | 53 | |

We have included two sets of observed numbers because the first set from at least three oppositions, though on average somewhat less accurate, includes more recent measures reaching fainter levels, suggesting that there is no obvious dependence on the sizes of KB bodies. Figure 4 presents much the same data in a different format by plotting the i's corresponding to the totals in columns 2 and 6. Together they establish the result that what might seem a likely initial range of i, i(o) < 8 deg, in no way matches the observed distribution. We extend this conclusion by repeating in Fig. 5 a comparison from two other mmrs, 1/2 and 2/5, though neither one, currently with 41 and 21 members with a's lying within +/-0.5 AU of the mmr's center, is nearly so well populated as 2/3. These two cases do have the advantage of integrations to 4.6 byr and the small coincidental aesthetic plus of having nearly equal numbers of observed and calculated objects. Here again, the observed KBOs show a far greater dispersion in the i's when the comparison group is defined by i(o) < 8 deg. Both the figures and Table I show an interesting feature, not just the fact that nearly half the bodies have i's > 10 deg, but correspondingly, that another half with i's < 10 deg, clearly does exist. So large a number we shall soon argue could readily be preserved from a broad range of earlier values but on the other hand their presence would question any process [analogous to migration] that acted to elevate the entire inclination range.

Horizontal comparisons, observed vs calculated, for the e's in Figs. 4 and 5 do not benefit from an ideal display, nor are they of great present interest. But we can infer from Fig. 4 a reasonable agreement of the two at 2/3 in the region, 0.1 < e < 0.3, and for 1/2 and 2/5 in Fig. 5, for 0.15 < e < 0.35. The poor agreement arising from a shortage of the calculated number outside these limits is just the consequence of a model that did not stock more bodies over a greater initial range in semimajor axis so that captures might have ended there after the migration ceased.

Mention of migration brings up the topic referred to earlier: we need to verify that it basically affects only the e's and cannot materially alter the initial inclination distribution when subject to a migration over 10 myr. Specific theoretical arguments claim this should be the case, but we also wished to address this matter by providing the numerical results presented in Table II. It lists two sets of average increases, di = [<i> - i(o)], where i(o) < 8 deg, for KBOs liberating at the 2/3, 1/2 and 2/5 mmrs and also a composite set of 52 that lie in 8 mmrs with semimajor axes between 3/5 and 1/3, lying between 42.3 and 62.6 AU. The entries in cols. 3 and 4 apply to a time shortly after the migration's effective end; those after 4.6 byr in cols. 5 and 6 are final, cumulative increases that include values in cols. 3 and 4. Data for 2/3 come from integrations extending to 1 byr. Included in Table II are di's corresponding to ~ the third of the total number at any mmr having the highest mean value. This choice allows an approximate comparison with the observations as Table I shows that about 1/3 of their total have i's elevated into the 10 - 20 deg level.

**Table II. Average increases in inclination, di, from initial values, i(o), at various resonances following a migration of 7 AU with a smooth e-folding time of 10 myr, whence cols. 3 and 4 give di's after 4 e-folding times. Final 2 columns show increases after 4.6 byr, though only for 1 byr at 2/3, hence the *. Initial values are randomly chosen for i(o) < 8 deg, e(o) < 0.1 and 29.0 < a(o) < 39.3 AU. All integrations include the 4 major planets, where the inner 3 move with their current 3d elements, but Neptune migrates out to its present semimajor axis at 30.09 AU maintaining its small, almost constant e and i.**

| mmr | number | di(40 myr) ( deg.) | di(largest 1/3) | di(f) | di (largest 1/3) (after 4.6 byr) | |
|---|---|---|---|---|---|---|
| 2/3 | 53 | 0.9 | 3.2 | 1.1* | 4.0* | |
| 1/2 | 38 | 1.2 | 4.7 | 2.7 | 6.9 | |
| 2/5 | 19 | 4.1 | 7.4 | 6.1 | 10.5 | |
| 3/5 – 1/3 | 52 | 2.5 | 6.3 | 3.8 | 8.7 | Total of 8 mmr |
| " | 33 | 1.5 | 3.7 | 1.9 | 5.4 | 7 mmr, skip 2/5 |

Table II moves toward confirming that migration can only very slightly affect the initial inclinations. A possible exception occurs for the 3rd order 2/5 mmr which alone among the others has about a third, 7 of 19, of its members showing an average increase, first to as much as 7.4 deg and later to 10.5.  The small number of only 7 examples that lasted to 4.6 byr at two other 3rd order mmrs, 4/7 and 5/8, also developed comparably large di's.  Evidence of this behavior at 2/5 is also apparent in Fig. 5.  Although remaining to 4.6 byr, all orbits in 2/5 and in 1/2 with average i's > 10 deg are markedly chaotic with Lyapunov times, measured in periods of Neptune, less than 2500. Over that time not one of the 19 bodies captured into 2/5 escaped, but nearly half (31) of the 69 into 1/2 did [cf Ref. 1].

 Apart from this one clearly documented case at 2/5, on the basis of Table II we rest the claim that the observed distribution of the i's in Table I and in Figs. 1, 3, 4 and 5 for all but a 3rd order mmr shows very little evidence of increases from migration, nor from long-term planetary perturbations and may therefore represent something very similar to the initial one.  Figures 4 and 5, especially the former, reinforce this result as they indicate, for ~ 0.1 < e < 0.35, that the number distribution of the i's [open symbols] is independent of e.

To make the suggestion that the current i range matches something close to a more primordial one means being certain that bodies can be captured in a mmr when i(o)'s are as large as ~ 30 deg.  Table III starts to answer this question by listing capture probabilities, P(c), for several i(o) ranges at 2/3 and 1/2.  The P(c)s apply to bodies remaining after 1 byr except for the second entry for 1/2 which obtains after 4.6 byr.  "Temporary" P(c)s at much earlier times are much higher as many escapes occur during and shortly after migration.  This behavior serves to deplete a region once mmrs pass through it.  All cases here are based upon the same set of initial elements save for the i(o)s, hence all have e(o) < 0.1.  Had we introduced an e range comparable to that of the i's, all P(c)s would have been substantially reduced.  The essential point of Table III is that increasing i(o) to 20-30 deg does show a reduction in P(c), and one that continues to drop markedly for i(o) up to 35 deg, yet captures at both 2/3 and 1/2, though resulting in a growing number of more chaotic orbits, still do occur.

**Table III.  Capture probabilities, P(c).  All values apply after 1 byr except the one at 1/2 from Ref. 1 which extends to 4.6 byr.  [After just 1 byr, P(c)s at 1/2 also lay at 27 %.]  Paucity of numbers at 1/2 in rows 1 to 3 arises thanks to a smaller collecting region for 1/2.  Examples employ an initial sample of 200 to 500 bodies.  [Here i(o) < 8 deg for the first 2 rows only].  All cases set e(o) < 0.1; no effort in this paper considers P(c)s for higher e(o) ranges.  Final column gives Lyapunov times after 1 byr in periods of Neptune, hence indicating the number of bodies with a good chance of surviving until today.**

| i range | 2/3 mmr | | 1/2 mmr | | No. in 2/3 |
| --- | --- | --- | --- | --- | --- |
| | P(c) in % | No. | P (c) in % | No. | with log T(L)>3.5 |
| 0 - 10 | 18 | 53 | 27 | 13 | 32 |
| 10 - 20 | 19 | 37 | 24 | 12 | 21 |
| 20 - 30 | 9 | 17 | ~7 | 4 | 8 |
| 30 - 35 | ~4 | 7 | | 2 | 2 |
| 0 – 10 * | -- | -- | 21 | 34 | -- |

* = From Ref. 1

Table III calls for an observational comparison. Recall the objects in a mmr with e < 0.1, plus a possible extension to 0.15 for better coverage, have undergone very little e increase during migration because they were initially located just shy of where 2/3 came to rest; these e's then lie very close to the e(o) values used for earlier tables. [As a side point: the reduced number at e < 0.1 in Fig.2 would follow from a smaller initial population or, more likely, the greater frequency of chaotic orbits with e up to 0.1, cf Ref. 1.] If the implied hypothesis that today's range in i is/was an early solar system feature, then the P(c)s in col.2 of Table III should, as they appear to, closely correspond to the observed numbers given below in Table IV.

**Table IV. Observed population at the 2/3 mmr for objects plotted in Figs. 1 and 2. These observed numbers are nicely consistent with the P(c)s in Table III and therefore also with the claim that the i distribution could be ~ primordial with very little change over time.**

| i range | no. for e < 0.1 | no. for e < 0.15 |
| --- | --- | --- |
| 0 - 10 | 10 | 29 |
| 10 - 20 | 11 | 30 |
| 20 - 30 | 5 | 10 |
| >30 | 2 | 5 |

Our conclusion argues that the inclination distribution in certain well-populated mean motion resonances of the Kuiper belt, evident today, is/was a characteristic of the very early solar system, either present initially, or, only shortly later, developed from some evolutionary mechanism. The sweeping of secular resonances might be one example as they provide a means of raising angular elements. But analyses of their effects on bodies in the asteroid belt have shown that, though e's are quickly and easily pumped, the i's are not.  Whether their scanning might have a different outcome in the Kuiper belt, or might be contingent on early planetary realignments suggested by the Nice model, is currently unknown.  The possibility that the i range is primordial has some interest as it suggests that the proto-disk generating the solar system may have had a warped or distorted structure at least, or especially, in its outer region.  Whatever dominates, developed or initial i's, the i's are now derivable only thanks to their distribution in resonances which fortunately, we can show, do carry information from a much earlier epoch.  The i's were probably once accompanied by a similarly broad range of eccentricities.  A number of these bodies suffered the fate of one, often several, captures followed by e increase and escape from mean motion resonances during or following migration.  They have formed a halo of unstable, temporary KB objects, some of which appear still to be present, or to have been replaced, at high e's and assorted i's.  This population has developed into what can truly and aptly be called a 'scattered disk'.  An extreme example is provided by (90377) Sedna with a = 532 AU, e = 0.857 and i = 11.9 deg -- extreme indeed, but of a type encountered a number of times in Ref. 1 after bodies had escaped, principally from 1/2.

These arguments lead to the view that the Kuiper belt is not so much a two component system, as it is a single one that has evolved from a common stock.  So, to answer the earlier query: Yes, the high i,e objects in Figs. 1 and 2 really do fit the label of defining 'the scattered disk', scattered because bodies were ejected from resonances often with e's large enough even to allow occasional close approaches to a major planet, but the captured i's, as argued here, may represent a very ancient distribution.

**Figure Captions**

Fig. 1. Inclinations, i, vs semimajor axis, a, for 712 well-observed Kuiper belt objects. Short horizontal lines are combined estimates of uncertainty in a as well as liberation amplitude. Any resonance for which its, or one of its, conjunctions with Neptune occur at its apocenter (e.g. the 7 listed at the top), a circumstance lending to stable capture, seems especially well populat4ed.

Fig. 2. e vs a for the same data set. The absence of bodies with e < 0.2 beyond ~ 48 AU strongly suggests that it, or more likely a few AUs less, defines the outer limit of asteroidal size bodies dating from the very early solar system. In Fig. 1, as here, nearly all bodies lying outside that limit lie in resonances: 7/15, 6/13, 5/11, 4/9, 3/7,..., having been captured at a(o)<a(1/2) and moved outward via migration. As we argue here, these bodies have largely retained their i's from a much earlier time and their presence beyond a(1/2) in no way compromises the conclusion [cf Ref. 2] that the initial population tapers off markedly after ~ 45 AU.

Fig. 3. Detailed plot of i's at the 2/3 mean motion resonance, a = 39.43 AU.

Fig. 4. Observed and calculated i vs e distribution at 2/3. Note that the i distribution shows no sign of dependence on e for 0.1 < e < 0.35.

Fig. 5. A repeat of Fig. 4 for the 1/2 and 2/5 resonances.

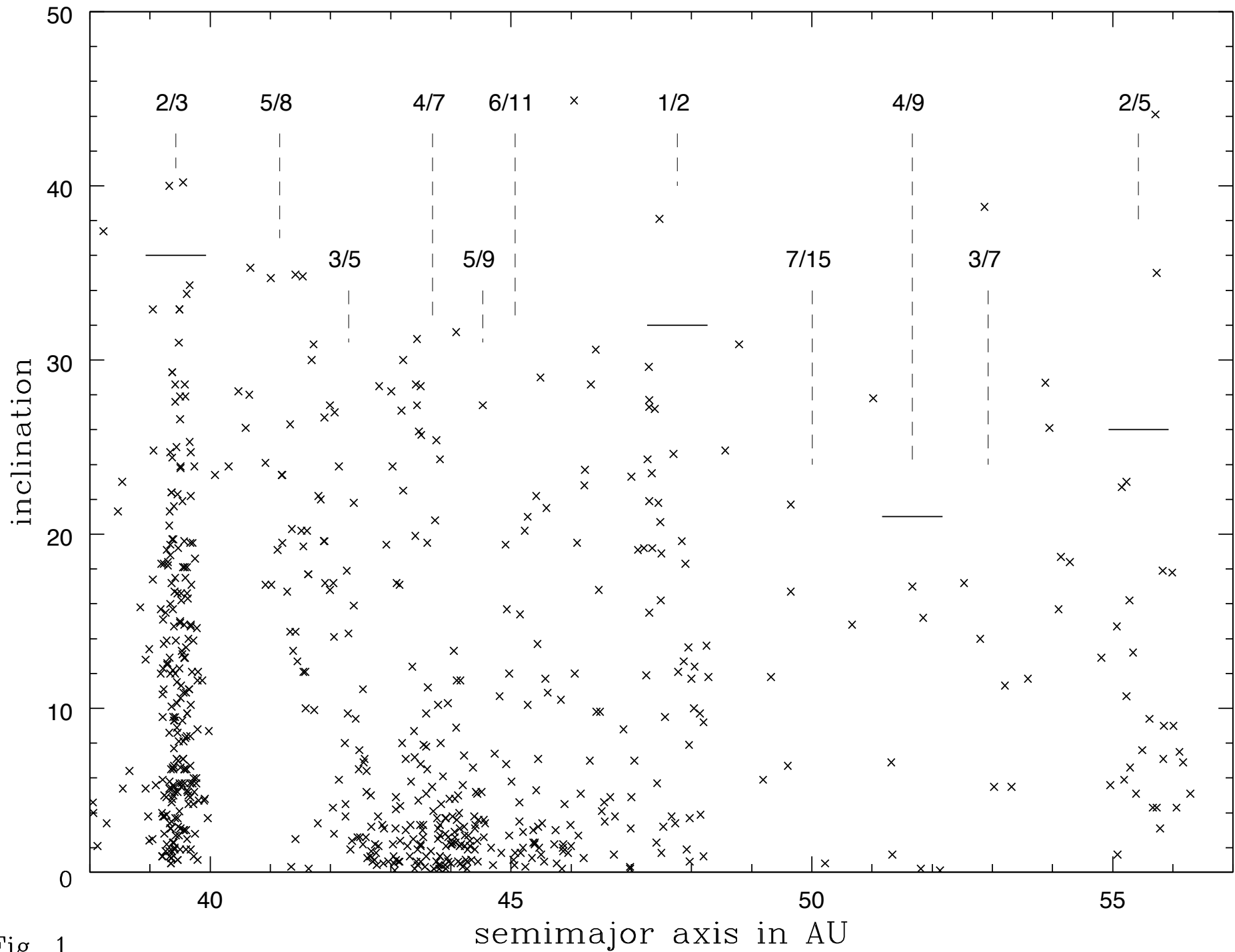

Fig. 1

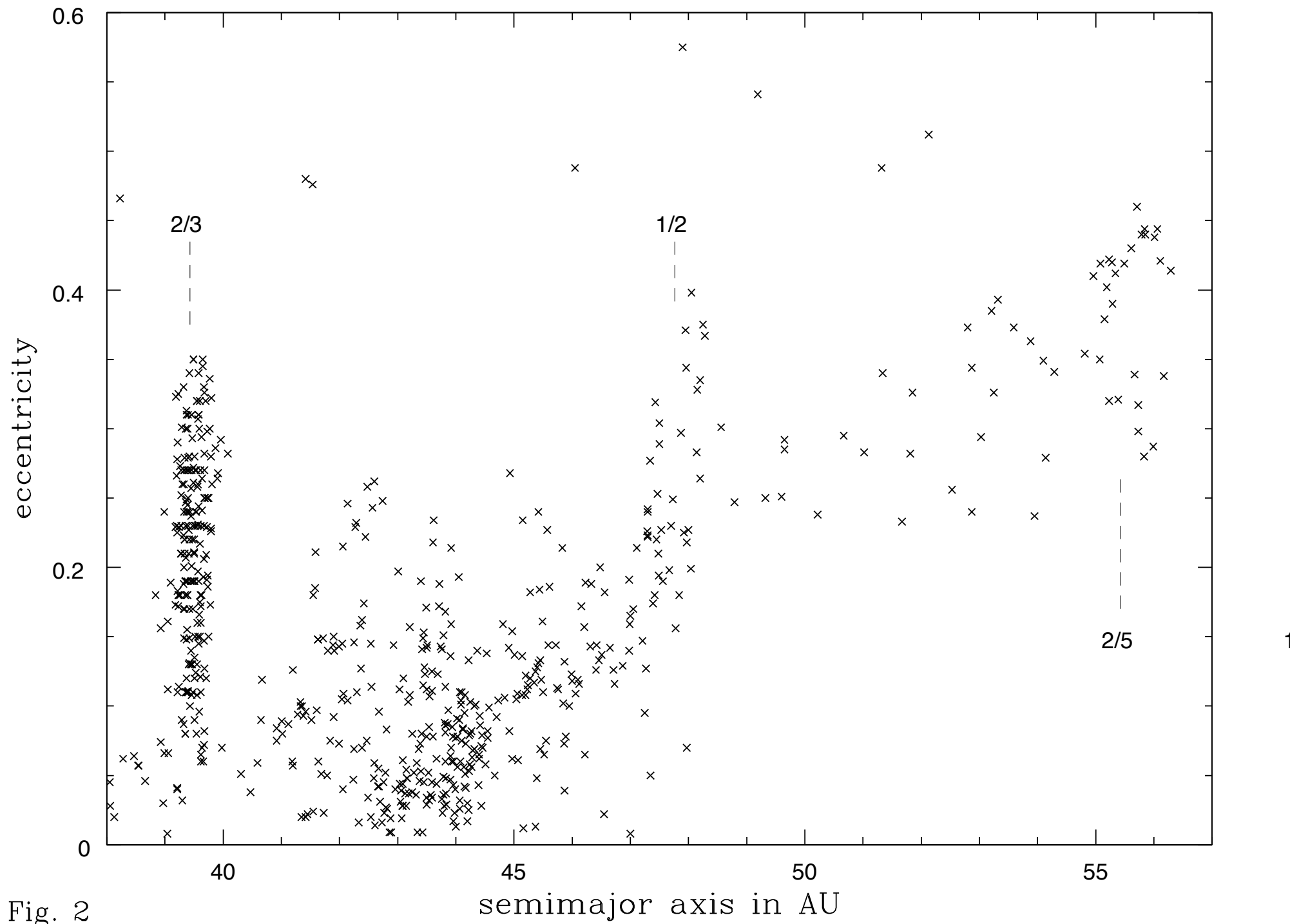

Fig. 2

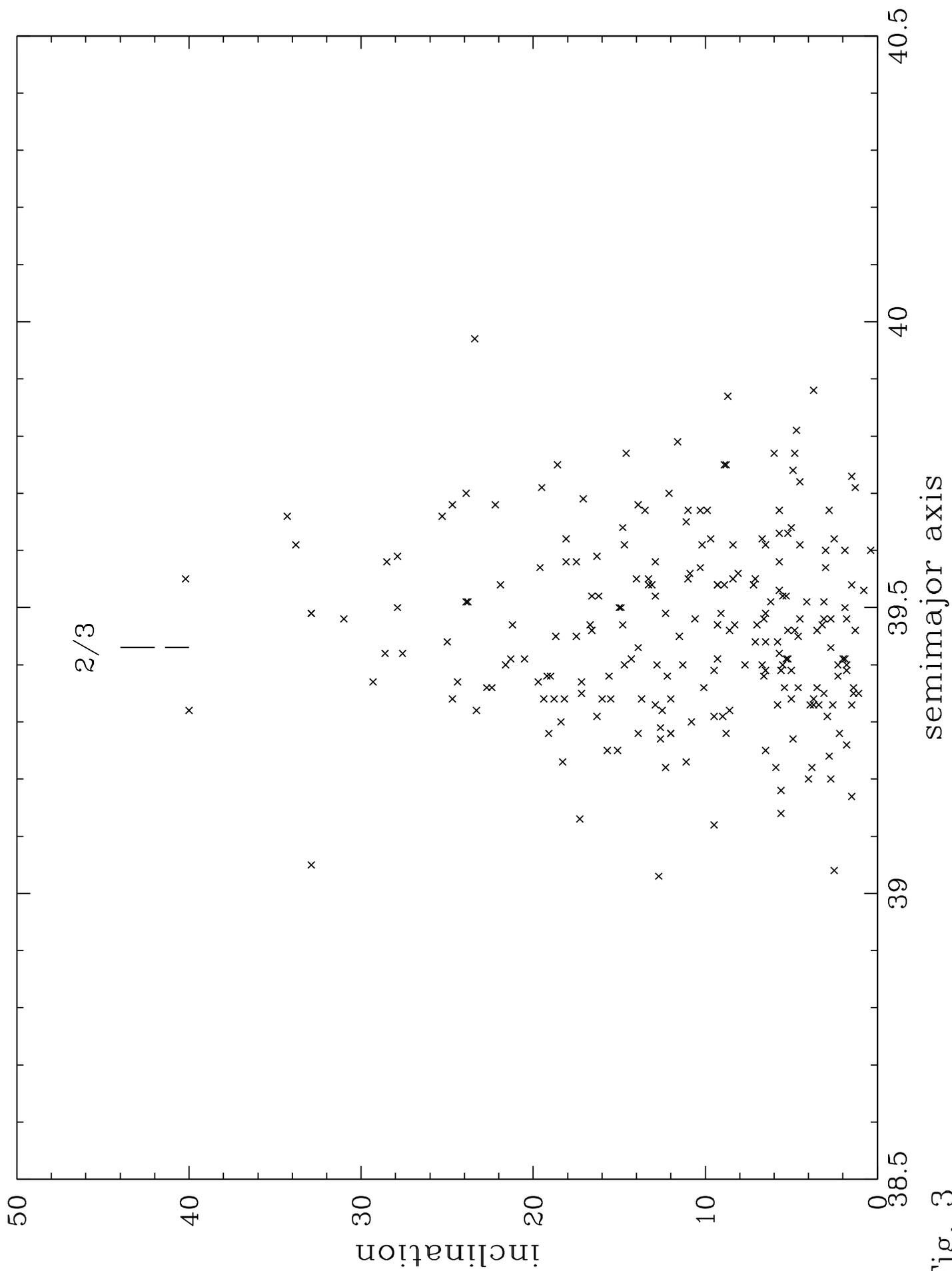

Fig. 3

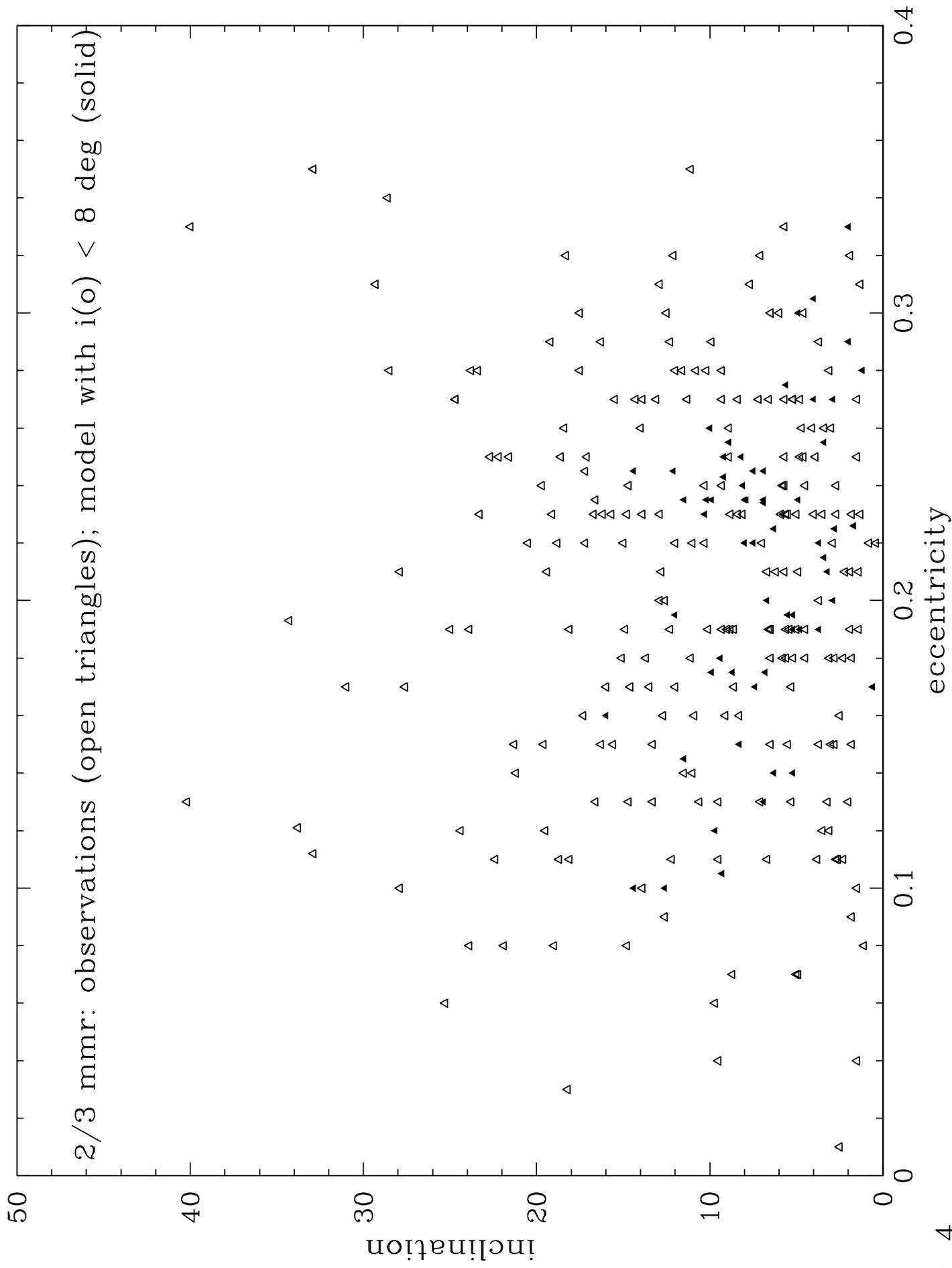

Fig. 4

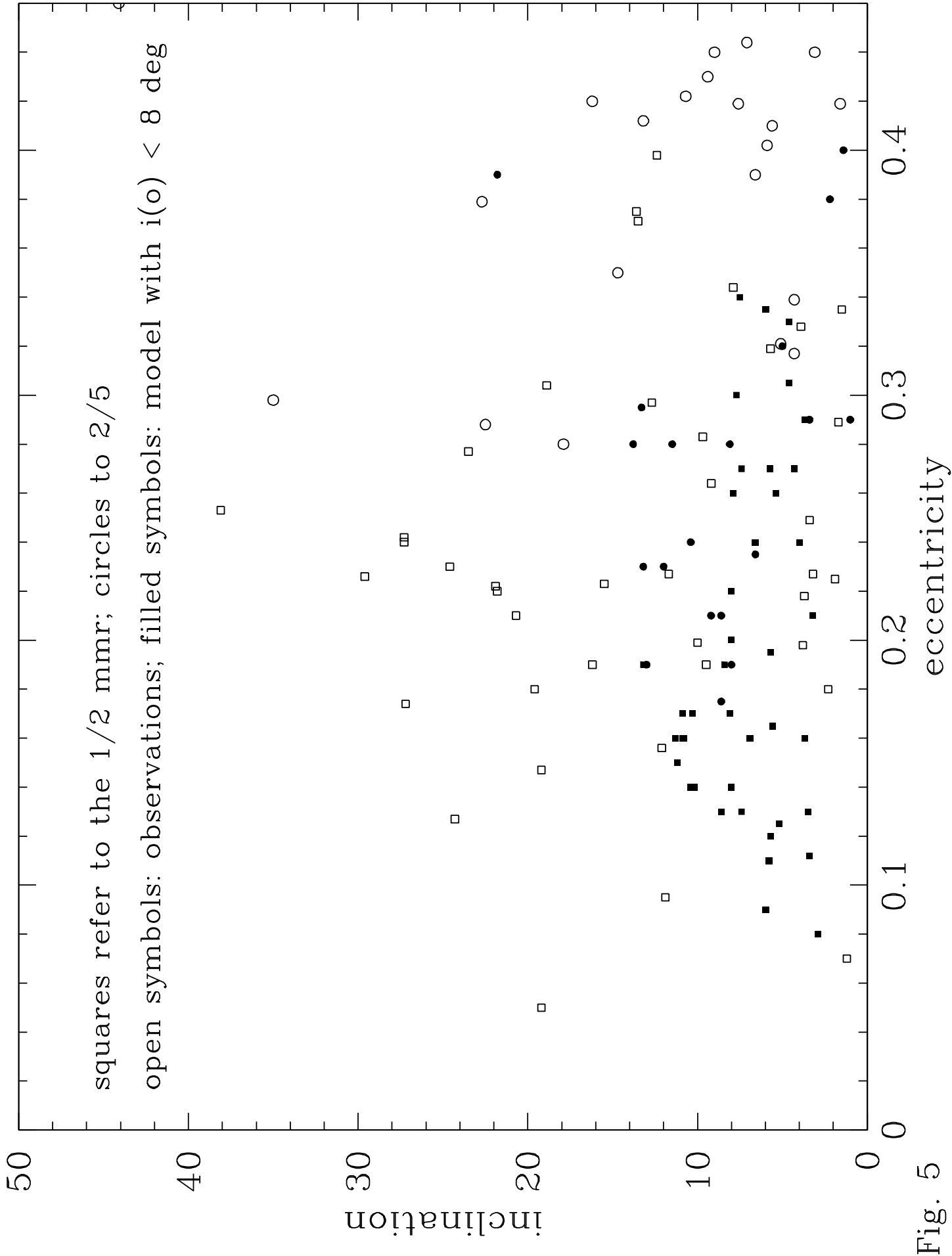

Fig. 5